\documentclass[aps,amsmath,amssymb,superscriptaddress,prb,reprint,twocolumn]{revtex4-2}
\usepackage{float}
\usepackage{dcolumn}% Align table columns on decimal point
\usepackage{multirow}
\usepackage{braket, bm}% bold math
\usepackage[utf8]{inputenc}
\usepackage[T1]{fontenc}
\usepackage{mathptmx}
\usepackage{tabularx}% table size
\usepackage[hidelinks]{hyperref}
\hypersetup{colorlinks, citecolor=blue, urlcolor=blue, linkcolor=blue}
\usepackage[dvipdfmx]{graphicx}

\usepackage{}	% required for `\align' (yatex added)

\begin{document}
\title{Wilson Loop and Topological Properties in 3D Woodpile Photonic Crystal}
\author{Huyen Thanh Phan} 
\email{phanthanhhuyenpth@gmail.com}
\affiliation{Department of Nanotechnology for Sustainable Energy, School
of Science and Technology, Kwansei Gakuin University, Gakuen Uegahara 1, Sanda 669-1330, Japan}

\author{Shun Takahashi}
\affiliation{Kyoto Institute of Technology, Matsugasaki, Sakyo-ku, Kyoto 606-8585, Japan}

\author{Satoshi Iwamoto}
\affiliation{Research Center for Advanced Science and Technology, The University of Tokyo, 4-6-1 Komaba, Meguro-ku, Tokyo 153-8505, Japan}

\author{Katsunori Wakabayashi}
\email{waka@kwansei.ac.jp}
\affiliation{Department of Nanotechnology for Sustainable Energy, School
of Science and Technology, Kwansei Gakuin University, Gakuen Uegahara 1, Sanda 669-1330, Japan}
\affiliation{Center for Spintronics Research Network (CSRN), Osaka University, Toyonaka 560-8531, Japan}
\affiliation{National Institute for Materials Science (NIMS), Namiki 1-1, Tsukuba 305-0044, Japan}

\date{\today}

\begin{abstract}
We numerically study the first and the second order topological states of electromagnetic (EM) wave in the three-dimensional (3D) woodpile photonic crystal (PhC). The recent studies on 3D PhCs have mainly focused on the observation of the topological states. Here, we not only focus on finding the topological states but also propose a numerical calculation method for topological invariants, which is based on the Wilson loop. For the 3D woodpile PhC, the topological states emerge due to the finite difference in the winding number or partial Chern number. The selection rule for the emergence of topological hinge states is also pointed out based on the topological invariants. Our numerical calculation results are essential and put a step toward the experimental realization of topological waveguide in 3D PhCs.
\end{abstract} 

\maketitle

\section{Introduction} \label{sec1}

Solid states physics has experienced a significant evolution since the topological phase of matter was introduced~\cite{Hasan2010,Qi2011,Ando2013,Bansil2016}. Topological insulators (TIs) are the materials derived from the application of topology to physics. They behave as insulators in the bulk region and as conductors at their boundaries. For example, 2D TIs offer topologically protected 1D edge states~\cite{Waka2009,Delplace2011,Yoshida2019,Obana2019, Song2020a, Koizumi2024}, while the 2D surface states are topologically protected in 3D systems~\cite{Moore2008, Yi2014, Benalcazar2017B,Ezawa2018,Schindler2018a, Notomo2023}.
Since the $d$-dimension topological systems give ($d-1$)-dimension boundary states, these states are called the first order topological states.
Recently, the studies of higher order topology have 
brought out the ($d-2$)-dimension topological protected boundary states ~\cite{Schindler2018a, Schindler2018, Langbehn2017, Song2017, Liu2019, Dutt2020}, which satisfy the conventional bulk-edge correspondence.
%extended the $(d-1)$-dimension topological edge states into lower dimension topological boundary states 

Inspired from the studies of topology in solid states physics, the topological phase is also found in PhCs~\cite{Haldane2008, Raghu2008, Chen2019, Liu2018, Ota2019, Phan2021, Vaidya2023, Wang2023} and sonic crystals~\cite{Lu2017, He2018, Xue2019, Zhang2019, Meng2020, Xue2019B, Ni2019}. In which, PhCs attract more attention because of their potential applications in telecommunication and devices~\cite{Joannopoulos1997, Butt2021, Mohammed2022}. The recent researches about topological PhCs are mainly focus on 2D systems, which are periodic in two dimensions and homogeneous in the third dimension. Higher order topological states in 2D PhCs are restricted in the second order, i.e. topological corner states localized in 1D corner~\cite{Ota2019, Phan2021, Phan2024}. Compared to 2D PhCs, 3D PhCs give one more dimension in the space to exploit topological properties. Besides, by using 3D PhCs, it is easier to guide the propagation of light in multi-directions. In principle, 3D PhCs can provide 2D surface states, 1D hinge states and zero-dimensional (0D) corner states, which are topologically protected by their crystalline structures. While surface and hinge states are applicable for signals transportation, 0D corner states are useful for optical cavities. 

To examine topological properties of PhCs, the topological invariant such as the Zak phase~\cite{Zak1989, Gao2015b, Wang2016, Blanco2020, He2022, Phan2024} is determined, which are often based on the parity of the wavefunction at high symmetric points~\cite{Liu2017, Kameda2019}. In the systems with inversion symmetry, the Zak phase is quantized to $\pi$ or $0$~\cite{Grusdt2013, Jiao2021}. On the other hand, when inversion symmetry is broken, the value of Zak phase is varied from $-\pi$ to $\pi$, which is called the Wilson loop. 
In our recent studies~\cite{Takahashi2024}, the Zak phase calculation method is explained for a 3D inversion symmetric PhC called simple cubic PhC, which is slightly different between the case for a single band and the case for a group of degenerate bands. To create the different topological properties in this photonic system, one of the simplest way is shifting the structure. For example, shifting the center point of the original unit cell by a half of the lattice constant leads to a new unit cell whose Zak phase is $\pi$ different from the Zak phase of the original one. 
The physics of these shifting structures can also be understood as the photonic "boundary-obstructed topological insulators"~\cite{Khalaf2021}, where the topological phases originate from the positioning of the boundaries.
We create the boundary between two types of unit cell and numerically obtained 2D topological interface states and 1D topological hinge states. However, the 1D hinge states in simple cubic PhC are found to be mixed with bulk and surface states. Therefore, EM wave propagating along the hinge leaks to bulk and surface very easily. Finding isolated hinge states in 3D PhC systems is essential for practical applications such as optical circuits or waveguides, since the energy-loss during the propagation process in this case is minimized.

In this paper, we theoretically design a 3D PhC structure following diamond cubic lattice and numerically study EM wave states in this structure. The topological states are numerically found between two types of unit cell which have the same photonic band structure but opposite topological properties. We also demonstrate a numerical method for the calculation of the Wilson loop on an arbitrary surface of 3D PhCs.
%without inversion symmetry. 
The emergence of 2D surface and the selection rule for 1D hinge states are explained based on the Wilson loop. Our numerical results of wave propagation in 3D woodpile PhC are essential and put a step toward the experimental realization of topological waveguide in 3D PhCs. 

\section{Woodpile Photonic Crystal}

\begin{figure*}[ht!]
    \centering\includegraphics[width=0.8\textwidth]{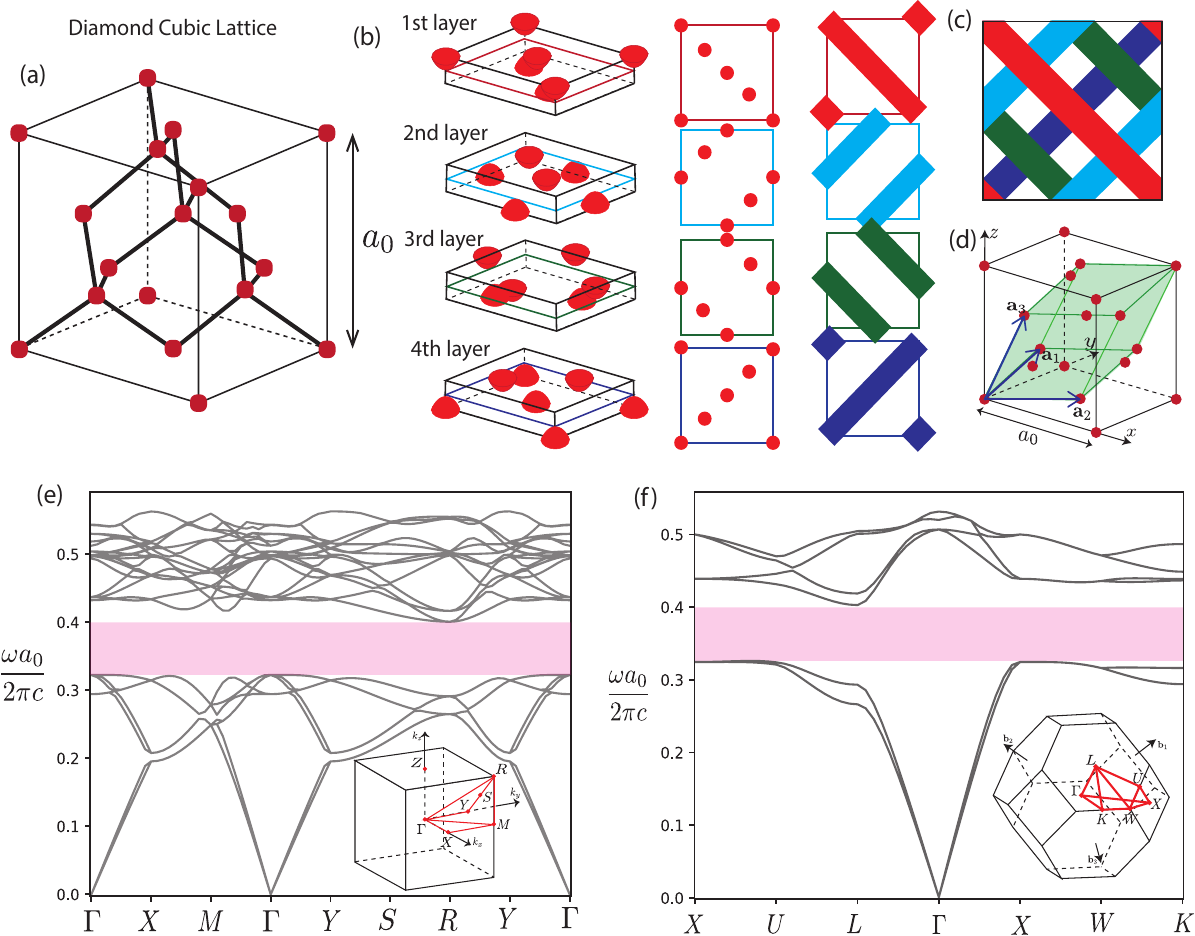}
    \caption{(a) A schematic of diamond cubic lattice. (b) Separated four layers of diamond cubic lattice which have the same thickness (left). The top view of each layer (middle). The arrangement of dielectric blocks in each layer, which follows the top view (right). Different colors indicate different layers. (c) A schematic of top view of conventional unit cell of woodpile PhC. (d) A schematic of primitive unit cell (green shaded region). ${\mathbf a}_1$, ${\mathbf a}_2$, ${\mathbf a}_3$ are 3 primitive lattice vectors of primitive unit cell. (e) Photonic band structure for conventional unit cell, the inset shows the corresponding first BZ. (f) Photonic band structure for primitive unit cell, the inset shows the corresponding first BZ.}
    \label{fig1}
\end{figure*}

We start from a woodpile lattice, which consists of dielectric blocks and air. Figure~\ref{fig1}(a) shows the schematic of a conventional unit cell of diamond cubic lattice. Lattice constant is $a_0$, the red dots represent for atomic sites.
The woodpile PhC is inspired from the diamond cubic lattice by dividing diamond cubic unit cell into four layers, which have the same thickness of $a_0/4$. For the atomic sites, which are cut through, each layer will contain a half of each site as shown in the left panels of Fig.~\ref{fig1}(b).
Then, by looking at the top view of each layer (middle panels), the dielectric blocks are arranged following the projected position of atoms onto layers (right panels). The different colors of blocks indicate different layers. The thickness of each block is $0.25a_0$, their width are $0.198a_0$. 
They are made of GaAs with the refractive index of $3.6$.

The conventional unit cell of the woodpile PhC is a cubic containing four layers. The top view is shown in Fig.~\ref{fig1}(c). By repeating this conventional unit cell infinitely in all $x$, $y$ and $z$ directions, we obtain the woodpile PhC. However, the diamond cubic lattice has a primitive unit cell (minimum unit cell), which is the parallelepiped labeled by green shaded region in Fig.~\ref{fig1}(d). Therefore, we can also define the primitive unit cell of woodpile PhC in the same way as primitive unit cell of diamond cubic lattice. Three primitive vectors of primitive unit cell are
${\mathbf a}_1 = \left(1/2, 0, 1/2\right)a_0$,
${\mathbf a}_2 = \left(1/2, 1/2, 0\right)a_0$,
${\mathbf a}_3 = \left(0, 1/2, 1/2\right)a_0$.
Their corresponding primitive vectors in the momentum space are 
${\mathbf b}_1 = 2\pi/a_0\left(1,-1, 1\right)$,
${\mathbf b}_2 = 2\pi/a_0\left(1, 1, -1\right)$,
${\mathbf b}_3 = 2\pi/a_0\left(-1, 1, 1\right)$.

Photonic band structure for the woodpile PhC is calculated by using finite-element method. Figure~\ref{fig1}(e) is the photonic band structure for conventional unit cell. The inset shows the corresponding first Brillouin zone (BZ), the red path indicate irreducible BZ. A complete band gap is observed around the normalized frequency 0.38. This band gap is relatively larger than the band gap in any other 3D structures made from the same material~\cite{Yablonovitch1993}. Below the band gap, there are eight photonic bands, which are connected by the degenerate points. Similarly, in Fig.~\ref{fig1}(f), we show the photonic band structure for the primitive unit cell. The inset is the first BZ, the red path is irreducible BZ. A complete band gap is also observed in the same frequency range as in Fig.~\ref{fig1}(e). Below this band gap, there are only two photonic bands, while there are eight bands in conventional unit cell. The reason for this difference is that the conventional unit cell is four times bigger than the primitive unit cell.

Although the conventional and primitive unit cells show the differences in their geometrical shape and photonic band structures, each of them has its own advantages in our study. For example, the cubic shape of the conventional unit cell makes it easier to construct interface, hinge or corner than the parallelepiped shape of the primitive unit cell. On the other hand, the intrinsic properties of woodpile PhC will not be changed when different shapes of unit cell are used for calculation. So, the primitive unit cell becomes more convenient for examining the topological properties of the band gap since it shows less number of photonic band below the band gap.

In Figs.~\ref{fig1}(e) and~\ref{fig1}(f), we display the photonic band structures calculated by using the conventional and primitive unit cells of woodpile PhC, which are derived directly from the diamond cubic lattice. As mentioned in section~\ref{sec1}, the shifting structure may lead to the finite difference in topological invariant. 
Therefore, in order to examine topological states in this woodpile structure, we also mention other unit cells which is derived by shifting the origin of the above conventional and primitive unit cells by $a_0/4$, $a_0/4$, $a_0/2$ in x, y and z directions, respectively.
Hereafter, we name them before-shifting conventional unit cell, before-shifting primitive unit cell, after-shifting conventional unit cell and after-shifting primitve unit cell. Since both before- and after-shifting unit cells represent the same lattice, their bulk band structures are identical. The difference between them will be studied in the next sections.

Before studying the topological properties, we note here that the woodpile PhC is in the presence of time-reversal symmetry and mirror symmetry. 
Besides, although the inversion symmetry is preserved in the diamond cubic lattice, it is broken in the woodpile PhC due to the arrangement of the blocks instead of atomic sites.
%Besides, although the mirror symmetry is still preserved, the inversion symmetry is broken in this structure. 
Therefore, the Zak phase of this PhC will not be quantized to $\pi$ and $0$. In this case, it will take the values varied from $-\pi$ to $\pi$, the set of all values on a plane in 3D systems is called the Wilson loop. By determining the evolution of Wilson loop, we can find out other topological invariants of 3D PhC.

\section{Wilson Loop in 3D Photonic Crystals} \label{ae}

\begin{figure*}[ht!]
    \centering\includegraphics[width=0.8\textwidth]{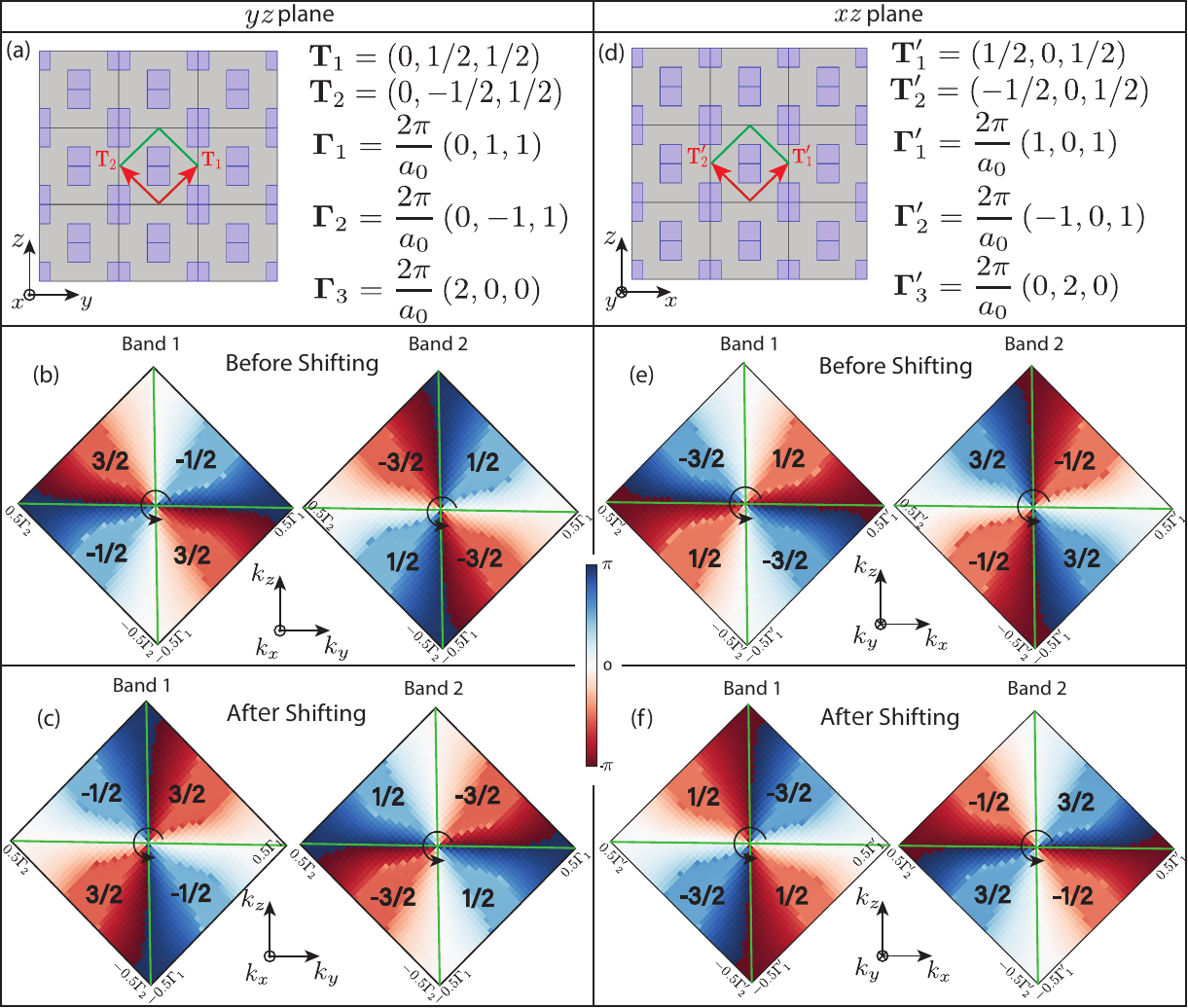}
    \caption{(a) The side view of woodpile PhC on $yz$ plane. ${\mathbf T}_1$ and ${\mathbf T}_2$ are two translational vectors of $yz$ plane, ${\bm \Gamma}_1, {\bm \Gamma}_2, {\bm \Gamma}_3$ are three reciprocal vectors used for calculation of the Wilson loop. The Wilson loop on $k_y k_z$ surface of before-shifting unit cell (b) and after-shifting unit cell (c). The numbers on each part of the Wilson loop results represent for the winding numbers. (d), (e), (f) are same as (a), (b), (c), but presented for $xz$ plane.}
    \label{fig2}
\end{figure*}
In this section, we will explain the numerical method for calculating the Wilson loop on arbitrary planes in 3D Woodpile PhC. Similar to the Zak phase, the Wilson loop is also defined for each photonic band. In Ref.~\cite{Takahashi2024}, we presented the numerical calculation method for Zak phase in 3D simple cubic PhC. It is slightly different between the case for a single band and for connected bands. While for a single band, we use directly the scalar products of periodic Bloch functions, for connected bands, we use the overlap matrices. 
For the woodpile PhC, if the conventional unit cell and its cubic first BZ are used to calculate the Wilson loop, the calculation formula and calculation steps will become similar to the simple cubic lattice, where the overlap matrices are used. However, in this case, because there are eight connected photonic bands below the complete band gap, each overlap matrix will be an $8 \times 8$ matrix. The numerical calculation will become complicated and time-consuming. Therefore, for more convenience, we will take advantages of the primitive unit cell, which contains only two photonic bands below the complete band gap, to calculate the Wilson loop.

Here we note that three primitive vectors of the primitive unit cell are not perpendicular to each other, their corresponding vectors in the reciprocal space are also not perpendicular to each other, which means that the first BZ is neither a cubic nor a cuboid. Therefore, the first BZ corresponding to the primitive unit cell cannot be simply used for calculation.
To generalize the Wilson loop calculation for 3D woodpile PhC to any arbitrary plane, the first BZ corresponding to the primitive unit cell should be replaced by a parallelepiped in the reciprocal space, which is formed by three vectors ${\bm \Gamma}_1$, ${\bm \Gamma}_2$, ${\bm \Gamma}_3$.
In particular, if the arbitrary 2D surface in the 3D woodpile PhC is assumed to be periodic under two translation vectors ${\mathbf T}_1$ and ${\mathbf T}_2$. Their corresponding vectors in reciprocal space are ${\bm \Gamma}_1$ and ${\bm \Gamma}_2$, respectively. The relationship between ${\mathbf T}$ and ${\bm \Gamma}$ is ${\mathbf T}_i \cdot {\bm \Gamma}_j = 2\pi \delta_{ij}$, where $i,j = 1,2$. The 2D surface that the Wilson loop depends on is the parallelepiped formed by ${\bm \Gamma}_1$ and ${\bm \Gamma}_2$.
The integration direction will be defined as a vector ${\bm \Gamma}_3$ that is orthogonal to the $({\bm \Gamma}_1, {\bm \Gamma}_2)$ plane and satisfies the following equation
\begin{equation}
    \label{newBZ}
    {\bm \Gamma}_3 \cdot ({\bm \Gamma}_1 \times {\bm \Gamma}_2) = {\mathbf b}_3 \cdot ({\mathbf b}_1 \times {\mathbf b}_2).
\end{equation}
Equation~(\ref{newBZ}) indicates that volume of the new momentum zone formed by three vectors ${\bm \Gamma}_1, {\bm \Gamma}_2, {\bm \Gamma}_3$ equals to the volume of the first BZ of the primitive unit cell.

For the arbitrary surface which is paralleled to the $({\bm \Gamma}_1, {\bm \Gamma}_2)$ surface, the Wilson loop for the isolated $n$-th band is determined by the following equation
\begin{equation}
    \label{wilsonloop}
    \mathcal{W}^n \left({\bm \Gamma_1}, {\bm \Gamma_2}\right) = \int_{-{\bm \Gamma}_3/2}^{{\bm \Gamma}_3/2} 
    \braket{{\mathbf u}^n_{\mathbf k} |i \partial_{\bm \Gamma_3}|{\mathbf u}^n_{\mathbf k}} d{\bm \Gamma_3}, 
\end{equation}
where ${\mathbf k} = {\bm \Gamma_1} + {\bm \Gamma_2} + {\bm \Gamma_3}$, ${\mathbf u}^n_{\mathbf k}$ is the periodic part of the Bloch wave functions which contains all three spatial components of both electric and magnetic field. For numerical calculation, we use the discrete limit of Eq.~(\ref{wilsonloop}) as follow
\begin{equation}
    \label{discrete_wilsonloop}
    \mathcal{W}^n \left(k_i, k_j\right) = 
    -\operatorname{Im} \left( \log \prod_{l} \braket{\mathbf{u}^n_{k_i ,k_j, k_l}|\mathbf{u}^n_{k_i, k_j, k_{l+1}}} \right),
\end{equation}
here, $k_i, k_j, k_l$ are the discrete $k$-points on ${\bm \Gamma_1}, {\bm \Gamma_2}, {\bm \Gamma_3}$, respectively.
For a group of $N$ degenerate bands, the scalar products in Eq.~(\ref{discrete_wilsonloop}) are replaced by the following overlap matrices 
\begin{equation} 
\label{omatrix}
S_{\mathbf{k}_1 \mathbf{k}_2} =
\begin{bmatrix}
\braket{\mathbf{u}^1_{\mathbf{k}_1} |\mathbf{u}^1_{\mathbf{k}_{2}} } & \braket{\mathbf{u}^1_{\mathbf{k}_{1}} |\mathbf{u}^2_{\mathbf{k}_{2}} } & ... &\braket{\mathbf{u}^1_{\mathbf{k}_{1}} |\mathbf{u}^N_{\mathbf{k}_{2}} } \\
\braket{\mathbf{u}^2_{\mathbf{k}_1} |\mathbf{u}^1_{\mathbf{k}_2} } & \braket{\mathbf{u}^2_{\mathbf{k}_1} |\mathbf{u}^2_{\mathbf{k}_2} } & ... &\braket{\mathbf{u}^2_{\mathbf{k}_1} |\mathbf{u}^N_{\mathbf{k}_2} }\\
... & ... & ... & ...\\
\braket{\mathbf{u}^N_{\mathbf{k}_1} |\mathbf{u}^1_{\mathbf{k}_2} } & \braket{\mathbf{u}^N_{\mathbf{k}_1} |\mathbf{u}^2_{\mathbf{k}_2} } & ... &\braket{\mathbf{u}^N_{\mathbf{k}_1} |\mathbf{u}^N_{\mathbf{k}_2} }
\end{bmatrix},
\end{equation}
where $\mathbf{k}_1 = (k_i,k_j,k_l)$, and $\mathbf{k}_2=(k_i,k_j,k_{l+1})$. The total Berry connection (in a matrix form) can be obtained by 
\begin{equation} 
\label{zak3}
    \hat{S}\left(k_i,k_j\right) = \prod_{l} S_{(k_i,k_j,k_l)(k_i,k_j,k_{l+1})}.
\end{equation}
To evaluate the Wilson loop for each band, we need the $n$-th eigenvalue $s^n$ of the above Berry connection matrix. 
Then, the Wilson loop for the $n$-th subband is given by 
\begin{equation} 
\label{zak4}
    \mathcal{W}^n \left(k_i, k_j\right) = -\operatorname{Im} \log
    \left(s^n\right).
\end{equation}

We note here that it does not like 2D PhCs, the polarization at $\Gamma$ point for 3D PhCs is not well-defined~\cite{Christensen2022}. Therefore, when doing the calculation for the Wilson loop, a special attention should be put on this point. There are several techniques to deal with this problem~\cite{Devescovi2024}. Since the Wilson loop is gauge invariant, in our work, we try to choose a gauge choice which avoid the $\Gamma$ point to calculate the Wilson loop.

Here, we apply this approach for the $yz$ and $xz$ surfaces in 3D woodpile PhC. In Fig.~\ref{fig2}(a), we show the side view of $yz$ surface, the periodicity of this surface is represented by two vectors $\displaystyle{{\mathbf T}_1 = (0, 1/2, 1/2)a_0}$ and $\displaystyle{{\mathbf T}_2 = (0, -1/2, 1/2)a_0}$. 
The corresponding vectors of ${\mathbf T}_1$ and ${\mathbf T}_2$ in reciprocal space are $\displaystyle{{\bm \Gamma}_1 = 2\pi/a_0\left(0,1,1\right)}$ and $\displaystyle{{\bm \Gamma}_2 = 2\pi/a_0\left(0,-1,1\right)}$. 
The third vector ${\bm \Gamma}_3$ is derived from Eq.~(\ref{newBZ}), which is $\displaystyle{{\bm \Gamma}_3 = 2\pi/a_0\left(2,0,0\right)}$.
Similarly, the side view of $xz$ plane is shown in Fig.~\ref{fig2}(d), together with the translational vectors ${\mathbf T}^{\prime}_1$, ${\mathbf T}^{\prime}_2$ and reciprocal vectors ${\bm \Gamma}^{\prime}_1$, ${\bm \Gamma}^{\prime}_2$, ${\bm \Gamma}^{\prime}_3$.

The Wilson loop on the $k_y k_z$ plane of the woodpile PhC will be calculated depending on the parallelogram formed by ${\bm \Gamma}_1$ and ${\bm \Gamma}_2$. The 1D integration will be taken along ${\bm \Gamma}_3$ direction.
Figures~\ref{fig2}(b) and~\ref{fig2}(c) are the Wilson loop on $k_y k_z$ surface for the two lowest bands of before-shifting and after-shifting unit cells, respectively.
Similarly, the Wilson loop on the $k_x k_z$ surface for two types of unit cell are displayed in Figs.~\ref{fig2}(e) and~\ref{fig2}(f).
Red and blue colors denote the negative values (from $-\pi$ to $0$) and positive values (from $0$ to $\pi$), respectively. For each plot of the Wilson loop, we divide the surface into four parts by the green lines and examine the evolution of Wilson loop in each part. Here, we define the winding number for the $n$-th band as follow
\begin{equation}
    \label{winding}
    w^n = \frac{1}{2\pi}\oint_{\text{L}} \partial_{\text{L}} \mathcal{W}^n(k_i, k_j) d\text{L}, 
\end{equation}
where $\text{L}$ denotes a closed loop on any surface in the first BZ. If $\text{L}$ is taken along the boundary of each part in the Wilson loop results, following the anti-clockwise direction, the winding number for each part is determined and written by the black half-integer numbers in Figs.~\ref{fig2}(b),~\ref{fig2}(c),~\ref{fig2}(e) and~\ref{fig2}(f).
In the topological crystal structures, Chern number is normally defined as the integration of Berry curvature over the first BZ~\cite{Fukui2005, Goudarzi2022}. On the other hand, Chern number is also known as being equal to the winding number of the Wilson loop in the first BZ~\cite{Wang_2019, Chen2021}. If the integration of Berry curvature is taken over a part of the first BZ, or the winding number of the Wilson loop is determined in a part of the first BZ, Chern number is re-named as partial Chern number.
Based on this analysis, the winding numbers indicated as the black half-integrer numbers in Figs.~\ref{fig2}(b),~\ref{fig2}(c),~\ref{fig2}(e) and~\ref{fig2}(f) equal to the partial Chern numbers of each corresponding part in the first BZ.
Comparing the before-shifting and after-shifting unit cells, it is easily seen that the partial Chern numbers are always different in all parts for both two bands and in both $yz$ and $xz$ surfaces. This topological difference may lead to the topological transition at the boundary between two types of unit cell. 

\section{Topological Interface States}

Owing to the partial Chern number difference in both $yz$ and $xz$ surfaces, the topological interface states are expected to form at the boundary between two types of unit cell. This will be numerically examined in this section.
\begin{figure}[h!]
    \centering\includegraphics[width=0.48\textwidth]{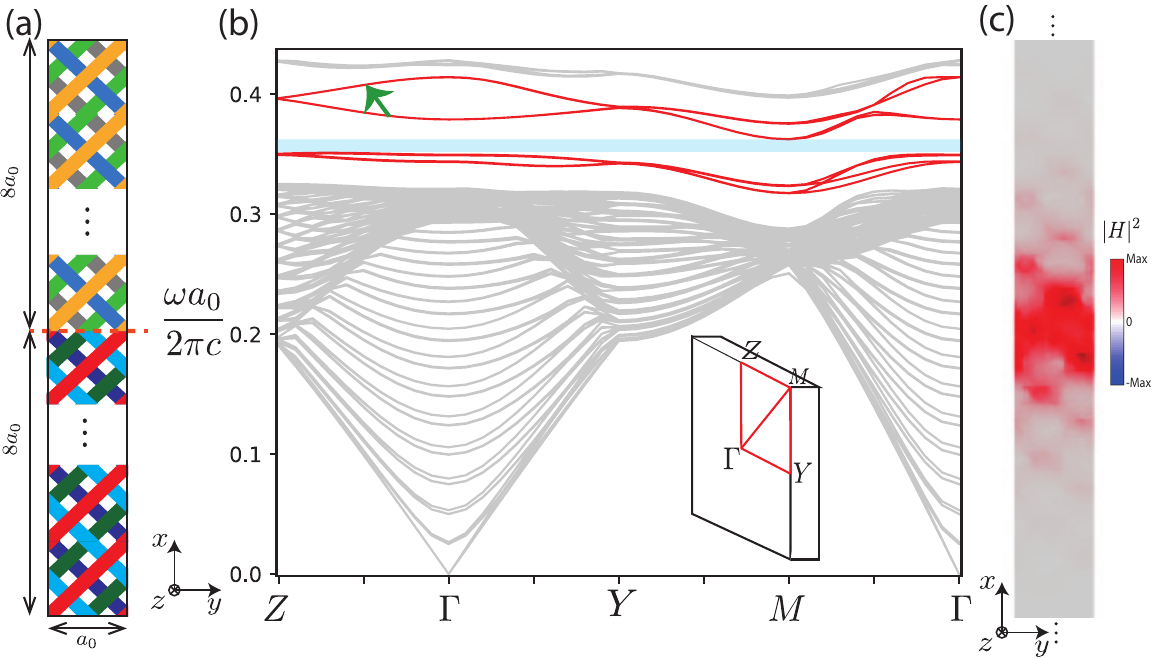}
    \caption{(a) A schematic of a supercell containing the interface parallel to $yz$ surface. This supercell contains eight before-shifting conventional unit cells (below the dashed line) and eight after-shifting conventional unit cell (above the dashed line). (b) Photonic band structure for the supercell in (a) with periodic boundary condition in three directions. The inset shows the first BZ for interface structure. Gray and red lines indicate bulk states and boundary states, respectively. (c) Field profile of the interface state denoted by the green arrow.}
    \label{fig3}
\end{figure}
Figure~\ref{fig3}(a) is a schematic of a supercell containing eight before-shifting conventional unit cells (below the dashed line) and eight after-shifting conventional unit cells (above the dashed line). The interface between them is parallel to $yz$ surface, which is denoted by the red dashed line. The corresponding first BZ of this supercell is shown in the inset of Fig.~\ref{fig3}(b). By applying periodic boundary condition in all three directions, we solve the eigenvalue equation and obtain photonic band structure for the supercell as shown in Fig.~\ref{fig3}(b). The gray lines indicate bulk states, the red lines present interface states where EM waves are localized at the boundary between two types of unit cell and decays exponentially. A field distribution at the interface state denoted by green arrow is shown in Fig.~\ref{fig3}(c). 

In the photonic band structure, we obtain eight interface states in the complete band gap of woodpile PhC. 
Between the fourth and the fifth interface state, we also see a complete band gap around the normalized frequency $0.36$, which is the shaded blue region in the Fig.~\ref{fig3}(b). 
The number of interface states can be explained by the partial Chern number difference of each band as shown in Figs.~\ref{fig2}(b) and~\ref{fig2}(c). 
In particular, the partial Chern number difference between before-shifting and after-shifting primitive unit cells is $\pm 2$ at all parts in the 2D BZ.
This is true for both two lowest bands.
Since these two connected bands have orthogonal polarization, their contribution to the topological properties at the band gap are independent.
Because the number of interface states equals to the partial Chern number difference~\cite{Hatsugai1993}, each band below the gap contributes two interface states, which means that four interface states should appear in the band gap as usual.
However, the area of the 2D BZ in the inset of Fig.~\ref{fig3}(b) (when using conventional unit cells) is a half smaller than the area of the 2D BZ in Figs.~\ref{fig2}(b),~\ref{fig2}(c) (when using primitive unit cells).
This folded in half of the BZ doubly increases the number of the interface states. Therefore, eight interface states are numerically observed in the band gap.
Along the high symmetric directions $\Gamma-Y$, $\Gamma-Z$ and $Y-M$, the woodpile PhC is mirror symmetric, interface states along these symmetric lines are doubly degenerate. On the other hand, along $\Gamma-M$ direction, the mirror symmetry is broken, leading to the non-degenerate interface states.

%Here we note that the interface states are doubly degenerate in almost $k$-points in high symmetric lines, except for $\Gamma-M$ region. This is because at the touching surface, both types of unit cell are mirror symmetric along $\Gamma-Y$, $\Gamma-Z$ and $Y-M$ lines, whereas it is broken along $\Gamma-M$ line.

By doing the similar manners, the interface structure which is paralleled to the $xz$ surface can also be constructed and examined. The numerical methods and results are identical to those used for $yz$ surface. Here we conclude that, the topological interface states are numerically confirmed to exist in 3D woodpile PhC due to the difference in partial Chern numbers between two types of unit cell. 

\section{Topological Hinge States}

Since the interface states (first-order) are numerically observed in both $yz$ and $xz$ surfaces, the topological hinge states (second-order) are also expected to occur at the boundary between them. In this section, we numerically examine 1D hinges formed by woodpile PhC, which is parallel to $z$ direction.
\begin{figure*}[ht!]
    \centering\includegraphics[width=0.8\textwidth]{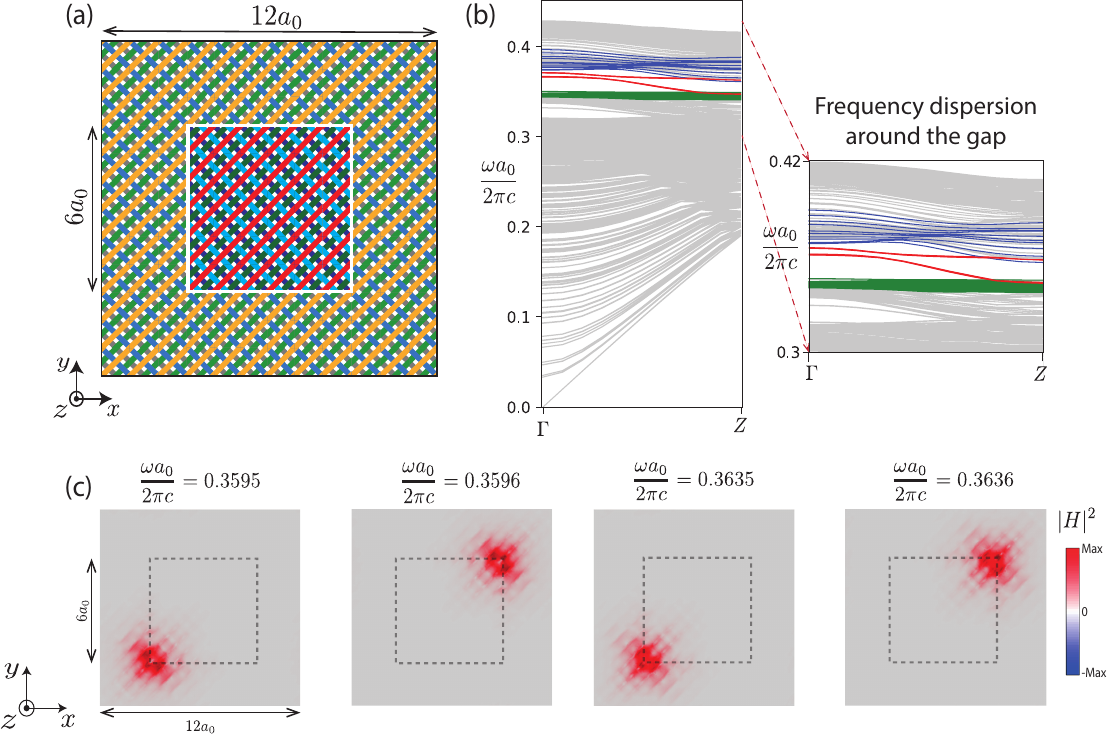}
    \caption{(a) A schematic of investigated supercell. The supercell size is $12a_0, 12a_0, a_0$ in $x$, $y$  and $z$ directions, respectively. At the center, $6\times 6$ before-shifting conventional unit cells are embedded, which is surrounded by 3 periods of after-shifting conventional unit cell. (b) Photonic band structure of the supercell in (a). Gray lines indicate bulk states, green and blue lines are interface states, red lines denotes hinge states. The inset shows the zoom in figure at around the band gap. (c) Magnetic field distribution of hinge states at ${\mathbf k} = (0,0,0.4)\pi/a_0$. The dashed lines indicate the boundary between two types of unit cell. The first two panels are degenerate and the last two panels are degenerate.}
    \label{fig4}
\end{figure*}
Figure~\ref{fig4}(a) is the schematic of investigated supercell, whose size is $12a_0, 12a_0, a_0$ in $x$, $y$  and $z$ directions, respectively.
At the center, $6\times 6$ before-shifting conventional unit cells are embedded, which is surrounded by three periods of after-shifting conventional unit cell. The boundary between them is indicated by the white line. This supercell contains four hinges, in which two opposite hinges are equivalent. We calculate photonic band structure for the supercell with periodic boundary condition in three directions by finite-element method and obtain the result as shown in Fig.~\ref{fig4}(b).
%Since there is a complete band gap in between the interface states in Fig.~\ref{fig3}(b), we also observe the band gap for the hinge structure around the normalized frequency 0.36.
The gray lines are bulk states, the interface states below the band gap are labeled by the green lines and the interface states above the band gap are indicated by the blue lines. In the gap between interface states, there are two isolated hinge states denoted by the red lines, each state is doubly degenerate.

Figure~\ref{fig4}(c) is the magnetic field distribution of the hinge states at ${\mathbf k} = (0,0,0.4)\pi/a_0$.
From these field profiles, we see that the double degeneracy of hinge states is resulted from the geometrical equivalence of the two opposite hinges in the supercell. However, hinge states only emerge at upper right and lower left 
positions 
and they are absent at other two
positions. 
Since these hinges are geometrically formed by the interfaces paralleled to $yz$ and $xz$ surfaces, the selection rule may relate to the Wilson loop on $k_y k_z$ and $k_x k_z$ planes shown in Figs.~\ref{fig2}(b),~\ref{fig2}(c),~\ref{fig2}(e) and~\ref{fig2}(f). 
Making a bridge to our previous study~\cite{Phan2021}, we reported about the topological edge and corner states in the 2D honeycomb PhC structures. Since the honeycomb PhCs exhibit the valley degree of freedom, the valley Chern numbers are obtained for each valley. The topological edge states are explained by the valley Chern number difference. The emergence topological corner states are resulted from the sign flip of the valley Chern number difference in two edges forming the corners.
Comparing the 2D honeycomb PhC and the 3D woodpile PhC, the partial Chern number in our 3D woodpile PhC is similar to the valley Chern number in 2D honeycomb PhC. The topological interface states in the woodpile PhC are due to the partial Chern number difference between two types of unit cells. The similar analysis can be applied to explain for the emergence or absence of the hinge states in woodpile PhC.

\begin{figure*}[ht!]
    \centering\includegraphics[width=0.8\textwidth]{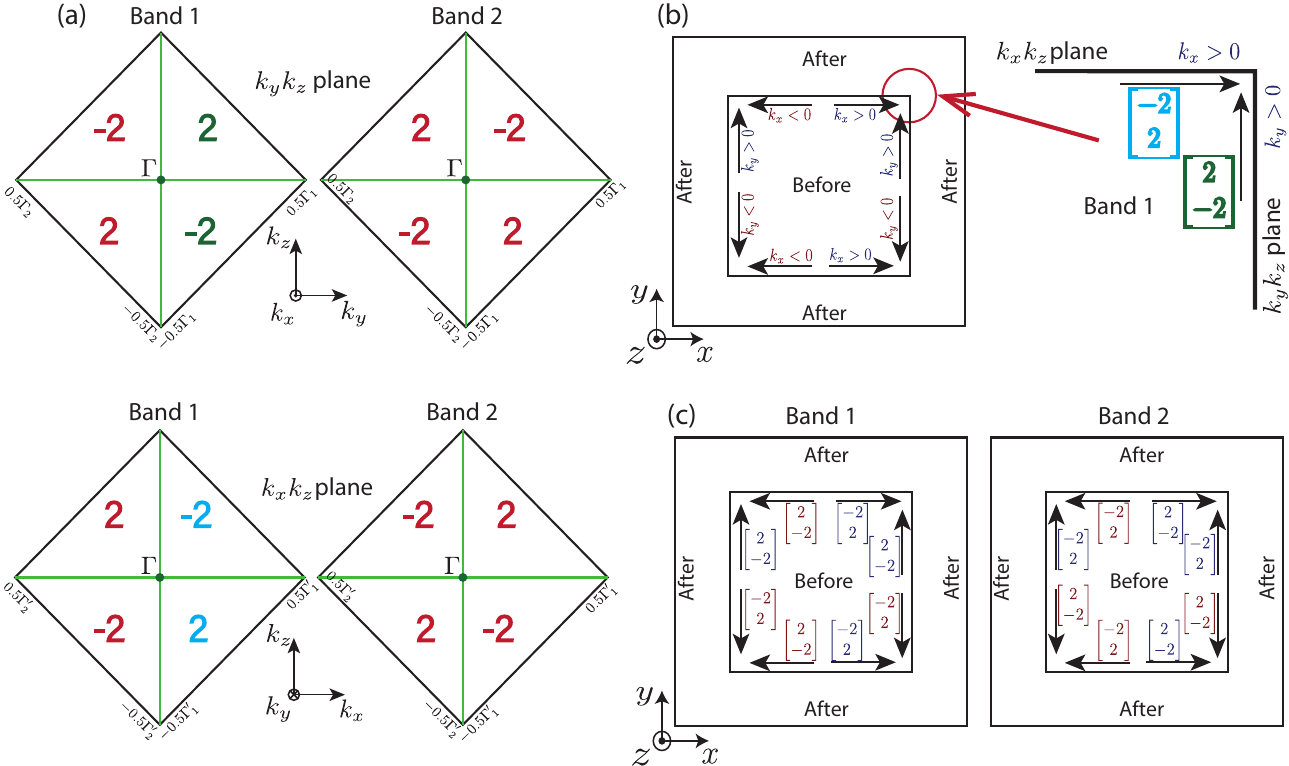}
    \caption{(a) The partial Chern number difference in each part of $k_yk_z$ and $k_xk_z$ surfaces for both two bands below the gap of woodpile PhC, calculated by using primitive unit cell. 
    (b) Schematic of EM waves propagating toward the hinges from  two interfaces. The inset shows an example of constructing the matrix $\begin{bmatrix} C^n_{+} & C^n_{-}\end{bmatrix}^T$ for the first band of the upper right hinge. The blue (green) matrix represents for $k_x k_z$ ($k_y k_z$) plane, this value is taken from the blue (green) numbers in (a).
    (c) The matrices $\begin{bmatrix} C^n_{+} & C^n_{-}\end{bmatrix}^T$ for each interface in case of band 1 (left panel) and band 2 (right panel).
    }
    \label{fig5}
\end{figure*}

In Fig.~\ref{fig5}(a), we demonstrate the partial Chern number difference between after-shifting and before-shifting unit cells at each part in $k_y k_z$  and $k_x k_z$ surfaces by subtracting partial Chern numbers of before-shifting unit cell from thoses of after-shifting unit cell. 
Before explaining for this selection rule, we assume that EM wave propagating forward or backward $x$ ($y$) direction will have positive or negative $k_x$ ($k_y$), respectively as shown in the left panel of Fig.~\ref{fig5}(b). The black arrows indicate the propagation direction. At each propagation direction on each surface, we create a $2\times 1$ matrix $\begin{bmatrix} C^n_{+} & C^n_{-}\end{bmatrix}^T$, where $C^n_{+}$ is the partial Chern number difference for $k_z>0$ and $C^n_{-}$ is the partial Chern number difference for $k_z<0$, $n$ is the band index. 
For example, at the upper right
hinge 
as depicted in the right panel of Fig.~\ref{fig5}(b), when the EM waves flow from two interfaces toward the hinge, $k_x$ is positive on the $xz$ surface and $k_y$ is also positive on the $yz$ surface. Therefore, $C^1_{+}=-2$, $C^1_{-}=2$ for $xz$ surface and $C^1_{+}=2$, $C^1_{-}=-2$ for $yz$ surface as written in the right panel of Fig.~\ref{fig5}(b). The blue (green) color of the matrix corresponds to the blue (green) numbers in Fig.~\ref{fig5}(a).
By doing the similar manners, the matrices for all interfaces are obtained as written in the Fig.~\ref{fig5}(c) for both two bands. It is easily seen that at the upper right and lower left
hinges, 
two matrices from two interfaces have opposite sign, whereas at the other two 
hinges, 
they are identical. The opposite sign of partial Chern numer difference is a typical topological transition signal, which causes the emergence of topological hinge states.

\section{Summary and Conclusion} 

We have theoretically designed a 3D Woodpile PhC following diamond cubic lattice and numerically studied the first and the second order topological states of EM wave in this structure by using finite-element method. The photonic band structure for before-shifting and after-shifting structures are identical, they show a relatively large complete band gap. Due to the broken inversion symmetry, the winding of the Wilson loop on a surface is used to determine the topological properties of this PhC. 
We have explained the numerical calculation method for the Wilson loop on an arbitrary plane in 3D PhC. By using this method, the finite difference in the partial Chern number between before-shifting and after-shifting structures are numerically confirmed. This topological difference causes topologically protected interface states at the boundaries between them.
When the before-shifting structure is surrounded by after-shifting structure, the interface and hinge states are observed at the 2D boundaries and 1D hinges. The important feature of these topological hinge states is that they emerge in the complete band gap between interface states and there is a selection rule for their emergence, which is based on the sign flip of the partial Chern numbers. 

The numerical calculation method of the Wilson loop in 3D PhCs is first-time introduced in this paper. This help us deeply understand about the topological properties of 3D PhCs and may helpful for the further prediction of other interesting phenomena may emerge in 3D PhCs.
Our numerical results of topological wave propagation in 3D woodpile PhC are essential and put a step toward the experimental realization of topological waveguide in 3D PhCs~\cite{PECS-Taka}. They are applicable in the communication technology.

\section*{Acknowledgements}

This work was supported by JSPS KAKENHI (Grants No. 22H05473, No. JP21H01019, and No. JP18H01154) , JST CREST (Grant No. JPMJCR19T1). K. W. acknowledges the financial support for Basic Science Research Projects  (Grants No. 2401203) from the Sumitomo Foundation.

\section*{Appendix A: Chern number and winding number in 3D systems}

Generally, in a 3D system, the Berry curvature is calculated by
\begin{equation}
    \begin{split}
        \mathbf{B}(\mathbf{k}) =& \; \nabla \times \mathbf{A}(\mathbf{k}) \\
        =& \;
        \begin{vmatrix}
            \hat{\mathbf{x}} & \hat{\mathbf{y}} & \hat{\mathbf{z}} \\
            \partial_x  & \partial_y  & \partial_z \\
            A_x & A_y & A_z
        \end{vmatrix} \\
        =& \; \hat{\mathbf{x}} \left(\partial_y A_z - \partial_z A_y\right) + \hat{\mathbf{y}} \left(\partial_z A_x - \partial_x A_z\right) + \hat{\mathbf{z}} \left(\partial_x A_y - \partial_y A_x\right) \\
        =& \; \hat{\mathbf{x}}B_x + \hat{\mathbf{y}}B_y + \hat{\mathbf{z}}B_z
    \end{split}
\end{equation}

\begin{figure}[h!]
    \centering\includegraphics[width=0.3\textwidth]{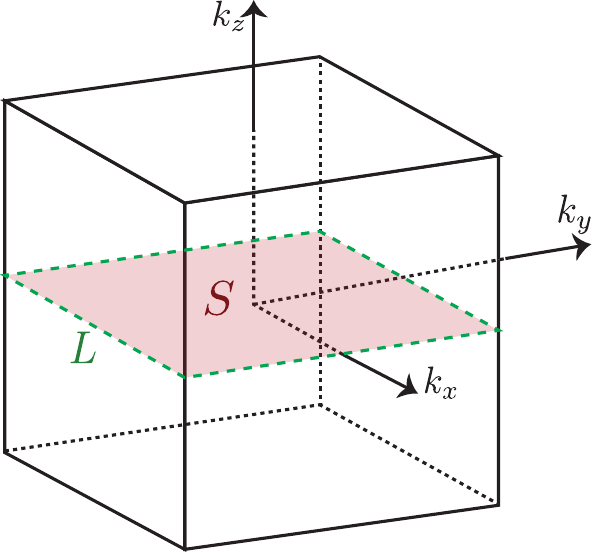}
    \caption{A momentum volume used to calculate the Wilson loop in a 3D system. The shaded surface $S$ is the surface where the Wilson loop is calculated. The green dashed line represent the loop $L$ surrounding the surface $S$.
    }
    \label{fig6}
\end{figure}

Here, we will explain in detail the calculation of Berry curvature $B_z$. The other two components $B_x$, $B_y$ can be calculated in a similar way. In Fig.~\ref{fig6}, we plot the first BZ of a 3D rectangular lattice, where the three basis vectors are paralleled to the $k_x, k_y, k_z$ directions, respectively. The shaded surface $S$ is the surface where the Berry curvature will be calculated. $L$ is the boundary loop of the surface $S$. The Berry curvature at each $k$-point on the surface $S$ is
\begin{equation}
\begin{split}
    B_z = \; \;&\partial_x A_y - \partial_y A_x \\
        = \; \;&{\mathbf \nabla}_{xy} \times \mathbf{A}(k_x, k_y, k_z).
\end{split}
\end{equation}
The total Berry curvature on the surface $S$ is
\begin{equation}
\label{1slice}
\begin{split}
    B_{zS} = &\int_{S} {\mathbf \nabla}_{xy} \times \mathbf{A}(k_x, k_y, k_z) dS \\
    = & \int_{L} \mathbf{A}(k_x, k_y, k_z) dL \\
    = & \int_{L} \langle {\mathbf u}_{L,k_z} | i\partial_L | {\mathbf u}_{L,k_z} \rangle dL.
\end{split}
\end{equation}

The Berry curvature in the surface $S$ in Eq.~(\ref{1slice}) is for one slice at a certain $k_z$ value. In the whole first BZ, the evolution of the Berry curvature in $k_z$ direction is given as
\begin{equation}\label{total}
\begin{split}
    B_{z}^{\Sigma} = &\int_{k_z} \partial_{k_z} \int_{L} \langle {\mathbf u}_{L,k_z} | i\partial_L | {\mathbf u}_{L,k_z} \rangle dL dk_z \\
    = & \int_{k_z} \partial_{k_z} \left[ \int_{L} \langle {\mathbf u}_{L,k_z} | i\partial_L | {\mathbf u}_{L,k_z} \rangle dL \right] dk_z \\
    = & \int_{k_z} \partial_{k_z} \mathcal{W}\left(k_z\right) dk_z.
\end{split}
\end{equation}

Thus, we obtain the following equation

\begin{equation}\label{total2}
\begin{split}
    \frac{1}{2\pi} B_{z}^{\Sigma} 
    = & \frac{1}{2\pi} \int_{k_z} \partial_{k_z} \mathcal{W}\left(k_z\right) dk_z \\
    = & \frac{1}{2\pi} \int_{L} \partial_{L} \mathcal{W}\left(L\right) dL,
\end{split}
\end{equation}
which indicates that the evolution of the Berry curvature in $k_z$ direction equals to the winding number of the Wilson loop along the boundary of each slice. This winding number also equals to the total Berry curvature of the four side surfaces or the Chern number~\cite{WindingChern}.  

% \nocite{*}
% \bibliography{apssamp}% Produces the bibliography via BibTeX. 

%\bibliographystyle{apsrev4-1}
\bibliography{references}
\end{document}